\documentclass[pra,onecolumn]{revtex4}%
\usepackage{amsfonts}
\usepackage{amsmath}
\usepackage{amssymb}
\usepackage{graphicx}%
\providecommand{\U}[1]{\protect\rule{.1in}{.1in}}

\begin{document}
\title{A large-scale one-way quantum computer in an array of coupled cavities}
\author{G. W. Lin$^{1}$}
\author{X. B. Zou$^{1}$}
\email{xbz@ustc.edu.cn}
\author{X. M. Lin$^{2}$}
\author{G. C. Guo$^{1}$}
\affiliation{$^{1}$Key Laboratory of Quantum Information, Department of Physics, University
of Science and Technology of China, Hefei 230026, People's Republic of China}
\affiliation{$^{2}$School of Physics and Optoelectronics Technology, Fujian Normal
University, Fuzhou 350007, People's Republic of China}

\begin{abstract}
We propose an efficient method to realize a large-scale one-way quantum
computer in a two-dimensional (2D) array of coupled cavities, based on
coherent displacements of an arbitrary state of cavity fields in a closed
phase space. Due to the nontrivial geometric phase shifts accumulating only
between the qubits in nearest-neighbor cavities, a large-scale 2D cluster
state can be created within a short time. We discuss the feasibility of our
method for scale solid-state quantum computation.

\end{abstract}

\maketitle

A quantum computer (QC) will exhibit advantages over its classical counterpart
only when a large number of qubits can be manipulated coherently, hence a
useful QC must allow control of large quantum systems, composed of thousands
or millions of qubits \cite{Shor}. Many architectures of QC's based on
scalable physical systems, such as ion traps \cite{Cirac,Kielpinski,Duan,Zhu},
optical lattices \cite{Jaksch}, semiconductor \cite{Makhlin}, have been widely
investigated. There are two well-known models for quantum computation, i.e.,
the quantum circuit model and measurement-based model. A class of
measurement-based models of quantum computation proposed by Raussendorf and
Briegel \cite{Raussendorf}, is the so-called cluster-state model, or one-way
quantum computer. Ref \cite{Raussendorf} has shown that two- and
three-dimensional (2D and 3D) cluster states can be used as universal resource
for quantum computation via local, single-qubit projective measurements and feedforward.

Recently, coupled cavity arrays
\cite{Helmer,Angelakis,Angelakis1,Irish,Huo,Zhou,Angelakis2,Hartmann,Rossini,Hartmann1,Hartmann2,Cho,Cho1}
have emerged as a fascinating alternative for simulating quantum many-body
phenomena and realizing quantum computing. In particular, theoretical works
have shown that the Mott-superfluid phase transition of polaritons
\cite{Hartmann,Rossini,Hartmann1,Angelakis2}, the Heisenberg spin chains
\cite{Hartmann2,Cho}, and fractional quantum hall state \cite{Cho1} can be
realized in the coupled cavity arrays. There are a variety of technologies
have been employed for realizing these systems, such as microwave circuit
cavities \cite{Wallraff,Rabl}, microtoroidal cavity arrays \cite{Armani,Aoki},
photonic crystal defects \cite{Bayindir}.

In this work, we propose a scaling method for one-way quantum computation with
spin-$\frac{1}{2}$ physical qubits in a 2D array of coupled cavities
\cite{Helmer,Cho1}. We find that when all the qubits are simultaneously
prepared in a spin state $\left\vert \downarrow\right\rangle $ or $\left\vert
\uparrow\right\rangle $, after coherent displacements of the quantum state
$\left\vert \Psi\right\rangle $ of cavity fields in a closed phase space, only
the qubits in nearest-neighbor cavities can fast accumulate a nontrivial
geometric phase shift, leading to creatation of a large 2D cluster state
within a very short time. Since the individual addressability of a qubit in
the coupled cavity array is easily performed, the 2D cluster state serves as
an effective resource for one-way quantum computation.

First we give a brief review of the geometric phase shift due to displacement
along an arbitrary path \cite{Leibfried,Zheng}. An arbitrary quantum state
$\left\vert \Psi\right\rangle $ of a harmonic oscillator can be coherently
displaced in the phase space. The effect of two sequential displacements
$D(\alpha)$ and $D(\beta)$ is additive up to a phase factor:%

\begin{equation}
D(\alpha)D(\beta)=D(\alpha+\beta)\exp[i\operatorname{Im}(\alpha\beta^{\ast})].
\end{equation}
For a path P consisting of N short straight sections $\Delta\alpha_{i}$,
$i=\{1,N\}$. The total operation is given by%

\begin{equation}
D_{total}=D(\Delta\alpha_{N})\cdot\cdot\cdot D(\Delta\alpha_{1})=\sum
\nolimits_{i=1}^{N}D(\Delta\alpha_{i})\exp\{i\operatorname{Im}[\sum
\nolimits_{i=2}^{N}\Delta\alpha_{i}(\sum\nolimits_{j=1}^{i-1}\Delta\alpha
_{j})^{\ast}]\}.
\end{equation}
Going to the limit of infinitesimal steps by replacing $\Delta\alpha_{i}$ with
$d\alpha$ yields:%

\begin{equation}
D_{total}=D[\int(d\alpha/dt)dt]\exp(i\gamma)\text{, with }\gamma
=\operatorname{Im}[\int\alpha^{\ast}(d\alpha/dt)dt].
\end{equation}
If the path P is closed, then:%

\begin{equation}
D_{total}=D(0)\exp(i\gamma)\text{, and }\gamma=\operatorname{Im}[\int
_{P}\alpha^{\ast}(d\alpha/dt)dt].
\end{equation}
The phase $\gamma$ is referred to as the geometric phase, which is independent
of the quantum state $\left\vert \Psi\right\rangle $. The fidelity of this
geometric phase gate due to the displacements, might be significantly higher
than that of the dynamical ones, as demonstrated in recent experiment in the
context of trapped ions \cite{Leibfried}. In the next part of this paper, we
will show that this geometric phase can be used for preparation of arbitrary
large 2D cluster state in principle. Finally, we discuss the feasibility of
our scheme. \begin{figure}[ptb]
\includegraphics[width=5.0in]{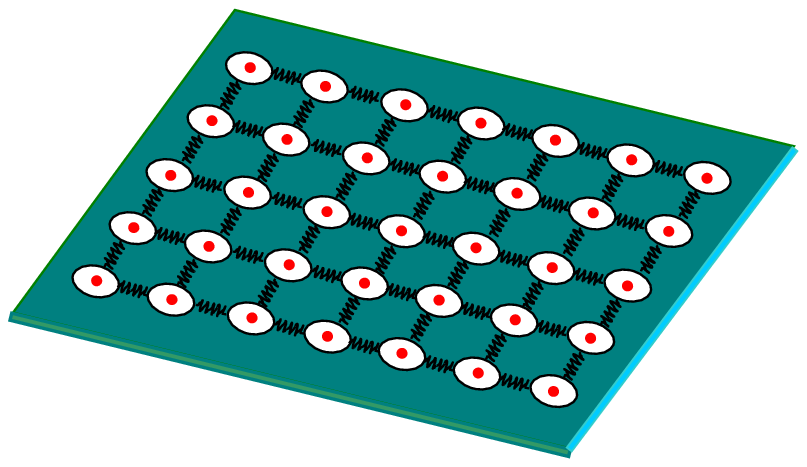}\newline\caption{(Color online)
Schematic representation of an array of coupled cavities. Each cavity traps a
spin-$\frac{1}{2}$ physical qubit, after coherent displacements of the quantum
state $\left\vert \Psi\right\rangle $ of cavity fields in a closed phase
space, a 2D square lattice cluster state can be achieved for universal quantum
computation via single-qubit measurements only.}%
\end{figure}

As sketched in Fig. 1, our model consists of a 2D ($M\times N$) array of
cavities that are coupled via exchange of photons with a spin-$\frac{1}{2}$
physical qubit in each cavity. The transition of two spin states $\left\vert
\uparrow\right\rangle _{m,n}$, $\left\vert \downarrow\right\rangle _{m,n}$ of
the qubit at the site $\{m,n\}$ couples to the cavity mode with the standard
Jaynes-Cummings type interaction $H_{I}=\sum\nolimits_{m=1}\sum\nolimits_{n=1}%
(a_{m,n}g_{m,n}\left\vert \uparrow\right\rangle _{m,n}\left\langle
\downarrow\right\vert +a_{m,n}^{\dagger}g_{m,n}\left\vert \downarrow
\right\rangle _{m,n}\left\langle \uparrow\right\vert )$, here $a_{m,n}%
^{\dagger}$ and $a_{m,n}$ are creation and annihilation operators for the
cavity mode at the site $\{m,n\}$, $g_{m,n}$ is the coupling strength. The
Hamiltonian that describes the photons in the cavity modes is $H_{cav}%
=\sum\nolimits_{m=1}\sum\nolimits_{n=1}\delta a_{m,n}^{\dagger}a_{m,n}%
+J\sum\nolimits_{m=1}\sum\nolimits_{n=1}(a_{m,n}a_{m,n+1}^{\dagger}%
+a_{m,n}a_{m+1,n}^{\dagger}+H.c.)$, where $\delta$ denotes the detuning of the
cavity mode from the transition of two spin states, $J$ is the tunneling rate
of photons. $H_{cav}$ can be diagonalized via the Fourier transform:
$a_{L,K}=\sqrt{MN}\sum\nolimits_{m=1}\sum\nolimits_{n=1}e^{i(Lm+Kn)}a_{m,n}$,
with $L=\frac{2\pi l}{M}$ and $K=\frac{2\pi k}{N}$ ($l=0,1,2,...M-1$,
$k=0,1,2,...N-1$), to give $H_{cav}^{^{\prime}}=\sum\nolimits_{L}%
\sum\nolimits_{K}\omega_{L,K}a_{L,K}^{\dagger}a_{L,K}$ with $\omega
_{L,K}=\delta+2J\cos L+2J\cos K$. Then $H_{I}$, switched to an interaction
picture with respect to $H_{cav}^{^{\prime}}$, can be rewritten as%

\begin{equation}
H_{I}^{^{\prime}}=\frac{1}{\sqrt{MN}}\sum\limits_{m=1}\sum\limits_{n=1}%
[\sum\limits_{L}\sum\limits_{K}ge^{-i(\omega_{L,K}t+Lm+Kn)}a_{L,K}\left\vert
\uparrow\right\rangle _{m,n}\left\langle \downarrow\right\vert +H.c.].
\end{equation}
\ \ 

Simultaneously we apply a classical field to each qubit, the interaction
Hamiltonian is described by $H_{cla}=\sum\nolimits_{m=1}\sum\nolimits_{n=1}%
\Omega_{m,n}\sigma_{m,n}^{x}$, where $\Omega_{m,n}$ is the Rabi frequency of
the classical field and $\sigma_{m,n}^{x}$ is the Pauli operator for the qubit
at the site $\{m,n\}$. In the strong driving regime $2\Omega_{m,n}\gg g_{m,n}%
$, $\omega_{L,K}$, we can realize a rotating-wave approximation and eliminate
the terms that oscillate with high frequencies, and obtain a new Hamiltonian
\cite{Solano}%

\begin{equation}
H_{I}^{^{\prime\prime}}=\frac{1}{\sqrt{MN}}\sum\limits_{m=1}\sum
\limits_{n=1}\sigma_{m,n}^{x}\{\sum\limits_{L}\sum\limits_{K}[ge^{-i(\omega
_{L,K}t+Lm+Kn)}a_{L,K}+ge^{i(\omega_{L,K}t+Lm+Kn)}a_{L,K}^{\dagger}]\},
\end{equation}
where we have assumed that $g_{m,n}=g$. The Hamiltonian in Eq. (6) is an
analogy to that for the high-speed gates with trapped ions simultaneously
interacting many vibrational modes \cite{Zheng}, except that the periodic
phase factor $e^{-i(Lm+Kn)}$ is dependent of the site for qubit in the array,
which is key importance, as shown below, for fast preparation of the cluster
states in parallel.

We define a new operator $J_{X}=\sum\nolimits_{m=1}\sum\nolimits_{n=1}%
[\sigma_{m,n}^{x}e^{i(Lm+Kn)}]$. The time-evolution operator for the
Hamiltonian in Eq. (6), based on Eq. (3), can be expressed as%

\begin{equation}
U(\tau)=\sum\limits_{L}\sum\limits_{K}[\exp(J_{X}\beta_{L,K}a_{L,K}^{\dagger
}-J_{X}^{\ast}\beta_{L,K}^{\ast}a_{L,K})\exp(i\gamma J_{X}^{\ast}J_{X})],
\end{equation}
with%

\begin{equation}
\beta_{L,K}=\frac{g}{\sqrt{MN}\omega_{L,K}}(1-e^{i\omega_{L,K}\tau}),
\end{equation}
and%

\begin{equation}
\gamma=\sum\limits_{L}\sum\limits_{K}\frac{2g^{2}}{MN\omega_{L,K}}[\tau
-\frac{\sin(\omega_{L,K}\tau)}{\omega_{L,K}}].
\end{equation}
Then we consider the state evolution under the operators in Eq. (7) and
$S_{z}=\prod\nolimits_{m,n}\sigma_{m,n}^{z}$ (i.e., single-qubit operation
$\sigma^{z}$ to each qubit ) by turns%

\begin{align}
U^{^{\prime}}(t)  &  =S_{z}U(\tau)S_{z}U(\tau)=\prod\nolimits_{m,n}%
\sigma_{m,n}^{z}\sum\limits_{L}\sum\limits_{K}[\exp(J_{X}\beta_{L,K}%
a_{L,K}^{\dagger}-J_{X}^{\ast}\beta_{L,K}^{\ast}a_{L,K})\exp(i\gamma
J_{X}^{\ast}J_{X})]\nonumber\\
&  \otimes\prod\nolimits_{m,n}\sigma_{m,n}^{z}\sum\limits_{L}\sum
\limits_{K}[\exp(J_{X}\beta_{L,K}a_{L,K}^{\dagger}-J_{X}^{\ast}\beta
_{L,K}^{\ast}a_{L,K})\exp(i\gamma J_{X}^{\ast}J_{X})]\nonumber\\
&  =\prod\nolimits_{m,n}\exp(i2\gamma J_{X}^{\ast}J_{X}),
\end{align}
where we have used the commutation relation $[S_{z},J_{X}^{\ast}J_{X}]=0$ and
anticommutation relations $\{S_{z},J_{X}\}=0$, $\{S_{z},J_{X}^{\ast}\}=0$. The
operator in Eq. (10) is equivalent to%

\begin{equation}
U^{^{\prime}}(t)=\exp[\sum\limits_{m^{^{\prime}}>m}\sum\limits_{n^{^{\prime}%
}>n}(i\Gamma\sigma_{m,n}^{x}\sigma_{m^{^{\prime}},n^{^{\prime}}}^{x})],
\end{equation}
up to an overall phase factor, where $\Gamma=4\gamma\cos[L(m^{^{\prime}%
}-m)+K(n^{^{\prime}}-n)]$. \begin{figure}[ptb]
\includegraphics[width=3.0in]{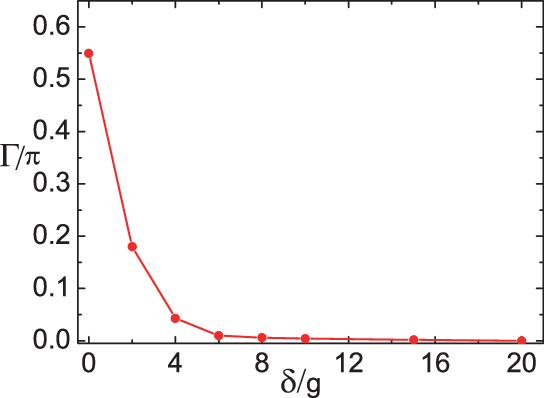}\newline\caption{(Color online) The
geometric phase shift $\Gamma$ as a function of $\delta/g$, assuming that
$M=N=19$, $J=0.1g$, $g\tau=3.$}%
\end{figure}\begin{figure}[ptb]
\includegraphics[width=5.0in]{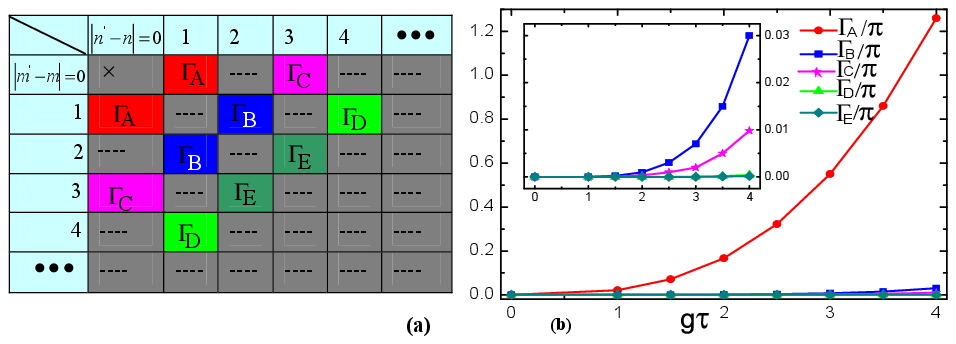}\newline\caption{(Color online)
Geometric phase shift $\Gamma$ versus the interaction time $\tau$ in units of
$1/g$, with different $\{\left\vert m^{^{\prime}}-m\right\vert ,\left\vert
n^{^{\prime}}-n\right\vert \}$. (a) The geometric phase shift $\Gamma_{\Theta
}$ ($\Theta=A$, $B$, $C$, $D$, $E$) with different $\{\left\vert m^{^{\prime}%
}-m\right\vert ,\left\vert n^{^{\prime}}-n\right\vert \}$. The symbol - - -
denotes that the geometric phase shift is smaller than $10^{-4}$. (b)
$\Gamma_{\Theta}$ versus the interaction time $\tau$ in units of $1/g$. Other
common parameters are $M=N=19$, $J=0.1g$, $\delta=0$.}%
\label{e}%
\end{figure}

The geometric phase shift $\Gamma$ has the following significant characters:
(i) when $\delta=0$, $\Gamma$ has a feasible value, while $\delta\gg g,J$,
$\Gamma\rightarrow0$. In Fig. 2, we plot $\Gamma$ a function of $\delta/g$,
from it we can see that when $\delta\gg10g$, $\Gamma$ is approximate to $0$.
(ii) Interestingly, if and only if $m^{^{\prime}}-m=\pm1$ and $n^{^{\prime}%
}-n=0$, or $m^{^{\prime}}-m=0$ and $n^{^{\prime}}-n=\pm1$, $\Gamma$ has a
feasible value, otherwise, $\Gamma\rightarrow0$, which is an analogy to
nearest-neighbor interaction in optical lattices \cite{Jaksch}. In Fig. 3, we
plot $\Gamma$ versus the interaction time $\tau$ in units of $1/g$, with
different $\{\left\vert m^{^{\prime}}-m\right\vert ,\left\vert n^{^{\prime}%
}-n\right\vert \}$. (iii) Another important result here is that $\Gamma$ is
independent of $M$ and $N$ \cite{444}, which means that the required time for
gate operation does not increase with the number of qubits.

In order to generate the cluster states, the initial state of all qubits in
coupled-cavity array should be prepared in the superposition of two
eigenstates of $\sigma_{m,n}^{x}$, for example the spin state $\left\vert
\uparrow\right\rangle $. The time evolution of the qubits under the the
operator in Eq. (11), when $4\Gamma=\pi$, the state of the qubits is
equivalent to a 2D cluster state. From Fig. 3, we see that the required time
for preparation is in the order of $1/g$ with $\Gamma=0.25\pi$. We note that
some theoretical schemes \cite{Helmer,Angelakis,Hartmann2} have proposed for
construction of 1D or 2D cluster states in coupled-cavity array. Besides
geometric phase shifts, our method, in principle, is suitable for preparation
of arbitrary large 2D cluster states in parallel, which provides the
possibility to implement scale quantum computation whin their coherence times.

Now we address the experiment feasibility of the proposed schemes. First, we
show that our method for solid-state qubit trapped in a 2D array of circuit
cavities \cite{Helmer}, in which solid-state qubits such as Cooper pair boxes
(CPB) and quantum dots (QD) are strongly coupled to circuit cavities
\cite{Wallraff,Rabl,Childress,Lin}, while the microwave photons have small
loss rates. The tunneling rate $J$ of photons between neighboring circuit
cavities has a feasible value about $100MHz$, and the qubit frequency can be
tuned in a large range. Typically, for CPBs interacting with the circuit
cavities \cite{Wallraff,Rabl}, the coupling strength is $g\sim2\pi\times
50MHz$, the photon lifetime is $T_{c}\sim1/\kappa_{c}\sim20\mu s$, and the
dephasing time of the two spin states $\left\vert \uparrow\right\rangle
_{m,n}$, $\left\vert \downarrow\right\rangle _{m,n}$ is $T_{a}\sim1us$. The
required time for preparation of arbitrary large-qubit cluster state, in
principle, is $T\sim0.01\mu s$ \cite{Here}, which is much shorter than $T_{c}%
$, $T_{a}$. For double-quantum-dot qubits trapped in the circuit cavities
\cite{Childress,Lin}, the coupling strength is $g^{^{\prime}}\sim2\pi
\times125MHz$, the photon decay time $T_{c}^{^{\prime}}\sim50\mu s$, the spin
dephasing time and charge relaxation time is about $T_{a}^{^{\prime}}\sim1\mu
s$. The required time for preparation of a large-qubit cluster state is
$T^{^{\prime}}\sim5ns\ll T_{c}^{^{\prime}}$, $T_{a}^{^{\prime}}$. For these
solid-state qubits, as shown above, the required time for preparation of
cluster state is smaller than microwave-photon coherence time, by about three
orders of magnitude. Therefore the cavity loss can be neglected in our
situation. The dephasing of the qubits themselves is the dominant source of
decoherence. After the solid-state qubits prepared in the cluster state, they
can be stored in the molecular ensembles \cite{Rabl}, which serve as a quantum
memory with a long coherence time. Second, for toroidal micro-cavities
\cite{Armani,Aoki}, in which the achievable parameters are predicted to be
$g^{^{\prime\prime}}\sim2.5\times10^{9}Hz$, spontaneous emission rate of the
high energy is $\kappa_{e}=1.6\times10^{7}Hz$, the photon decay time
$T_{c}^{^{\prime\prime}}\sim1/\kappa_{c}^{^{\prime}}\sim1/(0.4\times
10^{5}Hz)=25\mu s$, and the tunneling rate $J\sim1.6\times10^{6}Hz$. For
suppressing atomic spontaneous emission, two stable low levels, which are
coupled efficiently by a Raman process, are used as two spin states
$\left\vert \uparrow\right\rangle _{m,n}$, $\left\vert \downarrow\right\rangle
_{m,n}$. Thus the effective strength $g^{^{\prime\prime\prime}}\sim
1\times10^{8}Hz$, and the effective energy relaxation time $T_{a}%
^{^{^{\prime\prime}}}\sim100/\kappa_{e}\sim6\mu s$. The required time for
preparation of cluster state is $T^{^{\prime\prime}}\sim0.08\mu s\ll
T_{c}^{^{\prime\prime}},T_{a}^{^{^{\prime\prime}}}$.

In conclusion, we have provided an method to a large-scale one-way quantum
computer with spin-$\frac{1}{2}$ physical qubits in a 2D array of coupled
cavities. After coherent displacements of the quantum state of cavity fields
in a closed phase space, only the qubits in nearest-neighbor cavities can
accumulate a nontrivial geometric phase shift, which is key importance for our
scheme. We show the feasibility of our method for in various practical
systems. It seem that our scheme is most suitable for such solid-state system,
where the photons in the cavities have a long coherence time, effective
preparation of large-scale 2D cluster states can be achieved within a short time.

\textbf{Acknowledgments:} This work was funded by National Natural Science
Foundation of China (Grant No. 10574022 and Grant No. 60878059), the Natural
Science Foundation of Fujian Province of China (Grant No. 2007J0002), the
Foundation for Universities in Fujian Province (Grant No. 2007F5041), and
\textquotedblleft Hundreds of Talents \textquotedblright\ program of the
Chinese Academy of Sciences.


\textbf{References}

\begin{thebibliography}{99}                                                                                               %


\bibitem {Shor}P. Shor, in Proceedings of the 35th Annual Symposium on the
Foundation of Computer Science, edited by S. Goldwasser (IEEE Computer Society
Press, Los Alomitos, CA, 1994), pp.124--134.

\bibitem {Cirac}J. I. Cirac and P. Zoller, Nature 404, 579 (2000).

\bibitem {Kielpinski}D. Kielpinski, C. Monroe, and D. J. Wineland, Nature
(London) 417, 709 (2002).

\bibitem {Duan}L.-M. Duan, E. Demler, and M. D. Lukin, Phys. Rev. Lett. 91,
090402 (2003).

\bibitem {Zhu}S. L. Zhu, C. Monroe, and L. M. Duan, Phys. Rev. Lett. 97,
050505 (2006).

\bibitem {Jaksch}D. Jaksch, H.-J. Briegel, J. I. Cirac, C. W. Gardiner, and P.
Zoller, Phys. Rev. Lett. 82, 1975 (1999).

\bibitem {Makhlin}Y. Makhlin et al., Rev. Mod. Phys. 73, 357 (2001).

\bibitem {Raussendorf}R. Raussendorf and H. J. Briegel, Phys. Rev. Lett. 86,
5189 (2001).

\bibitem {Helmer}F. Helmer, M. Mariantoni, A. G. Fowler, J. Delft, E. Solano,
F. Marquardt, Euro. Phys. Lett. 95, 50007 (2009).

\bibitem {Angelakis2}D. G. Angelakis, M. F. Santos, and S. Bose, Phys. Rev. A
76, 031805(R) (2007). 

\bibitem {Angelakis}D. G. Angelakis and A. Kay, New J. Phys. 10, 023012 (2008).

\bibitem {Angelakis1}D. G. Angelakis, M. F. Santos, V. Yannopapas, and A.
Ekert, Phys. Lett. A. 362, 377 (2007).

\bibitem {Irish}E. K. Irish, C. D. Ogden, and M. S. Kim, Phys. Rev. A 77,
033801 (2008).

\bibitem {Huo}M. X. Huo, Ying Li, Z. Song, and C. P. Sun, Phys. Rev. A 77,
022103 (2008)

\bibitem {Zhou}Lan Zhou, Z. R. Gong, Yu-xi Liu, C. P. Sun, and Franco Nori,
Phys. Rev. Lett. 101, 100501 (2008).

\bibitem {Hartmann}M. J. Hartmann, F. G. S. L. Brand\u{a}o, and M. B. Plenio,
Nature Phys. 2, 849 (2006); A. D. Greentree et al., Nature Phys. 2, 856 (2006).

\bibitem {Rossini}D. Rossini and R. Fazio, Phys. Rev. Lett. 99, 186401 (2007).

\bibitem {Hartmann1}M. J Hartmann, M. B. Plenio, Phys. Rev. Lett. 99, 103601 (2007).

\bibitem {Hartmann2}M. J. Hartmann, F. G. S. L. Brand\u{a}o, and M. B. Plenio,
Phys. Rev. Lett. 99, 160501 (2007).

\bibitem {Cho}J. Cho, Dimitris G. Angelakis, S. Bose, Phys. Rev. A 78, 062338
(2008).  

\bibitem {Cho1}J. Cho, D. G. Angelakis, and S. Bose, Phys. Rev. Lett. 101,
246809 (2008).

\bibitem {Wallraff}A. Wallraff et al., Nature 431, 162 (2004).

\bibitem {Rabl}P. Rabl, D. DeMille, J. M. Doyle, M. D. Lukin, R. J.
Schoelkopf, and P. Zoller, Phys. Rev. Lett. 97, 033003 (2006).

\bibitem {Armani}D. K. Armani et al., Nature (London) 421, 925 (2003).

\bibitem {Aoki}T. Aoki et al., Nature (London) 443, 671 (2006).

\bibitem {Bayindir}M. Bayindir, B. Temelkuran and E. Ozbay. Phys. Rev. Lett.
84, 2140 (2000).

\bibitem {Leibfried}D. Leibfried, B. DeMarco, V. Meyer, D. Lucas, M. Barrett,
J. Britton, W. M. Itano, B. Jelenkovic, C. Langer, T. Rosenband, and D. J.
Wineland, Nature 422, 412 (2003).

\bibitem {Zheng}S. B. Zheng, Phys. Rev. A 74, 032322 (2006).

\bibitem {Solano}E. Solano, G. S. Agarwal, H. Walther, Phys. Rev. Lett. 90,
027903 (2003).

\bibitem {444}When $\delta=0$, $\Gamma$ has a maximal value, for $\omega
_{L,K}\neq0$, one can choose M(N) is odd.

\bibitem {Childress}L. Childress, A. S. S\O rensen, and M. D. Lukin, Phys.
Rev. A 69, 042302 (2004).

\bibitem {Lin}Z. R. Lin, G. P. Guo, T. Tu, F. Y. Zhu, and G. C. Guo, Phys.
Rev. Lett. 101, 230501 (2008).

\bibitem {Here}Here we have ignored the single-qubit operation times, since
which are much shorter than that for multiqubit operations.
\end{thebibliography}
\end{document}